\documentclass[a4,12pt]{article}
\begin{document}

        \hspace*{10cm}  PRA-HEP-98/6  \\

\begin{center}
{\Large \bf Are quantum teleportation and cryptography predicted by
     quantum mechanics? }  \\  [3mm]
{ Milo\v{s} Lokaj\'{\i}\v{c}ek }  \\
   {Institute of Physics, AV\v{C}R, 18221 Prague 8, Czech Republic}  \\
\end{center}

\begin{abstract}
It has been shown that the predictions of some new phenomena (e.g.,
teleportation and cryptography) are based on some assumptions added to the
quantum-mechanical model or modifying some of its basic axioms. The hitherto
experiments presented as a support of the mentioned phenomena may be hardly
regarded as sufficient, as they may be interpreted alternatively on the
basis of simple interference processes.  \\ [3mm]
\end{abstract}

      Some new phenomena are supposed to exist in the nature  (see, e.g.,
   \cite{tel,bou,cry}). It is argued that they are
   predicted on the basis of the current quantum-mechanical mathematical
   model and of its orthodox interpretation. However, in the given
  arguments some additional assumptions have been joined to the QM
  model, some of them raising  questions whether they are actually
consistent with other axioms the QM model is based on. In the following
we would like to call the attention to this fact.

    According to the orthodox interpretation it is necessary to
distinguish between the pure and mixed states: a pure state of a physical
system should change into a mixed state during a measurement, or rather in
any interaction with another object. A new special pure state  of microscopic
objects may arise, of course, in such an interaction. However, if two
different objects do not mutually interact and do not form a pure
state they represent a mixture being described with the help of a density
matrix; such a state can be hardly represented by one vector in a
Hilbert space (defined, e.g., as a direct tensor product of two simple
Hilbert spaces representing individual objects) and interpreted as a
pure state.

    And any predictions of teleportation or cryptography processes start from
the assumption that these  basic characteristics of the
quantum-mechanical model need not be necessarily fulfilled, or that
it is possible to pass freely from a mixed space to a pure space
(or vice versa?). It has been assumed in all corresponding (teleportation)
experiments that an entangled pair of photons has been prepared
with the help of down conversion in a non-linear crystal since the
two arising photons must be combined in a similar way as the photon
pair in EPR experiments. However, the proper EPR photons are assumed to
move then in a vacuum without any interaction  (between the source
and the detector) while the photons formed in the down conversion
must go through the rest of a crystal, which can hardly occur
without a series of secondary interactions inside the given
macroscopic object; the properties of the photon being definitely
settled in the surface plane of the crystal.  Therefore, according
to the standard quantum mechanics the considered photon pairs
(after leaving the crystal) should necessarily represent  mixed states.

    Another assumption consists in the possibility of interpreting the
system of two photons (e.g., the photons in Ref. \cite{bou} denoted as
Nr. 1 and 2, i.e.,  a mixture of two fully independent photons) as a pure
state in a Hilbert space spanned on four basic Bell's states although they
should be described with the help of a corresponding (general)
density matrix, like in the former case.

    Therefore, both these assumptions should be regarded as going
 behind  the standard quantum-mechanical model in which only  pure states
 may be
represented by single vectors in a Hilbert space; the model
proposed and formulated by J. von Neumann \cite{neu} in 1932. The
question arises then how to interpret the experiments being described
in the papers quoted in the beginning of this paper.  It is necessary
to ask whether these experiments can really contribute to a
justification of the mentioned modified assumptions, which has been
argued quite explicitly, e.g., in Ref.  \cite{bou}.

   We are afraid  that such arguments are very premature as there is
not any direct link between the given measurements and the mentioned
predictions. The measured characteristics may be explained in a
more natural way, e.g., as  simple and natural correlation of
processes being denoted generally as interference phenomena.  Any
new assumptions are not necessary; it is not necessary to relate
these phenomena to entangled states of photon pairs, either.

    The same arguments concern the very recent experiment performed by
 Tittel et al. \cite{tit}; see also Ref. \cite{tap}. Photons being formed
 by crystal down-conversion and going through very long optical fibers
 interact many times and the considered photon pairs represent again
state mixtures and not  pure states; it is a kind of decoherence (see,
    e.g. \cite{leg}).  In fact, we do not know practically anything
    what is going with photons passing through a crystal or optical
    fibers; it is known, however, that any observable change occurs
    practically always in the surface level of a given macroscopic
    object.

     And thus, the fiber lengths must be considered fully irrelevant
(with the exception
    of absorption) as the properties of any photon seem to be fixed in the
   instant when it leaves the last  object before going into
    the analyzer. All the mentioned  experiments indicate
    that a periodical characteristic  relates closely to some internal
  properties  of spinning photons, with the period corresponding to the
  given wave length.  The instantaneous phase of this characteristic
  influences then decisively the effect of a photon in an analyzer. The
  periodical property of the photon may be then
related to  a kind of the quantum phase considered recently, e.g.,
in Ref. \cite{schl}.

    It is evident that all the mentioned experiments may be interpreted on
the same basis as Newton fringes.  It concerns also the so
called entanglement swapping \cite{pan}, where a mixed state is
interpreted as a pure state without any actual substantiation (on the
basis of an interference measurement only). There is not any
experiment at the present
 time that would be able to distinguish between pure or mixed states under
the given conditions. Also the values of coincidence counts
obtained by combining different measuring channels may be
interpreted as a direct consequence of the just mentioned
characteristics.

    The hitherto experiments may be, therefore, hardly regarded as an actual
support for the existence of the announced teleportation and
cryptography phenomena. There is, however, an additional reason why
it is necessary to be very careful in doing similar predictions or
conclusions.  All such predictions are based on the existence of
entangled states of EPR type, which follows from the Copenhagen
interpretation.  And there are some other experiments which should
raise serious doubts, as they are questioning  the validity  of the
quantum-mechanical model itself, especially  its Copenhagen
version.

    These experiments have consisted in the measurement of photon (light)
   passage through three polarizers  \cite{kra1,kra2}. The transmittance
   dependencies on relative  angles between polarizer axes have contradicted
    convincingly the predictions
    based on Jones' matrices  which represent a generalization of QM
    description in the case of real (imperfect) polarizers. Consequently,
   other models concerning the description of microscopic phenomena should be
  tried and tested; different types of hidden variables  being included.

    Such a conclusion might seem, of course, to be in an important
    contradiction to the fact that Bell inequalities have been violated
   in the  standard EPR experiments (see, e.g.,  \cite{asp}).  However, it
has been shown by us quite recently that these inequalities do not
cover all possible kinds of hidden variables. An additional
(seemingly self-evident) condition has been included in all approaches
of their derivations, which excludes, e.g., the influence of
different impact points of photons into a microscopic grid of a
macroscopic device (polarizer). Bell's inequalities cannot be derived
    if one admits that an exact impact point together with the photon
    spin influence  results of measurement  \cite{lok}. They may be
derived only if at least one of these two factors is neglected.

   In addition to the just mentioned facts another important
    question concerns the general decoherence problem, when an
interaction with quantum vacuum is assumed to exist.  If such a
decoherence plays a role (as discussed in the last years)  one must ask
how long a pure state remains as a pure state.  In such a case it is
necessary to expect that the range of entangled state should be
strongly limited.  Additional doubts should, therefore, appear:
whether all involved assumptions may represent a consistent basis for
the given predictions.

   Summarizing all these aspects and arguments we must conclude that
  mutually contradicting assumptions seem to be applied to in the
  contemporary discussions of new phenomena (involving the EPR problem)
and that it is necessary to analyze newly not only these questions
alone but also practically all fundamentals of the
quantum-mechanical model, especially,  if some additional
assumptions going beyond this model are involved.

   Numerous discussions with my colleagues and friends J. Kr\'{a}sa and V.
Kundr\'{a}t about all related problems are highly appreciated.   \\


{\footnotesize
\begin{thebibliography}{99}
\bibitem{tel}
Ch. H. Bennett et al.: Teleporting an unknown quantum state via dual
classical and Einstein-Podolsky-Rosen channels; Phys. Rev. Lett. 70
(1993), 1895-9.
\bibitem{bou}
D. Bouwmeester et al.:  Experimental quantum teleportation; Nature
390 (1997), 575-9.
\bibitem{cry}
W. Tittel, G.Ribordy, N.Gisin:
Quantum cryptography; Physics World, March 1998, 41-5.
\bibitem{neu}
J. von Neumann: Mathematische Grundlagen der Quantenmechanik;
Springer - Berlin  1932.
\bibitem{tit}
W. Tittel, J.Brendel, B.Gisin, T.Herzog, H.Zbinden, N.Gisin:: Experimental
demonstration of quantum correlations over more than 10 km; Phys. Rev. A
57 (1998), 3229-32.
\bibitem{leg}
 A. Leggett: Time's arrow and the quantum measurement problem; in
"Time's arrow today (ed. S.F.Savitt), Cambridge Univ. Press 1995, pp.
97-106.
 \bibitem{tap}
 P. R. Tapster, J.G.Rarity, P.C.M.Owens:
 Violation of Bell's inequality over 4 km of optical fiber; Phys.  Rev.
 Lett.  73 (1994), 1923-6.
 \bibitem{schl}
 M.T. Fontenelle, M.Freyberger, H.Heni,
W.P.Schleich, M.S.Zubairy:  Quantum phase, photon counting and EPR
variables; The Dilemma of Einstein, Podolsky and Rosen - 60 Years
Later (eds. A. Mann.  M.Revzen), Inst. of Publish. Techno House,
Bristol 1996, pp. 73-82.
 \bibitem{pan}
 J.-W. Pan, D.Bouwmeester,
 H.Weinfurter, A.Zeilinger: Experimental entanglement swapping:
Entangling photons that never interacted; Phys. Rev. Lett. 80 (1998),
3891-4.
 \bibitem{kra1} J. Kr\'{a}sa, J. Ji\v{r}i\v{c}ka,
M.Lokaj\'{\i}\v{c}ek:  Depolarization of the light by imperfect
polarizers; Phys. Rev. E 48 (1993), 3184-6.
 \bibitem{kra2}
 J. Kr\'{a}sa, M.Lokaj\'{\i}\v{c}ek, J.  Ji\v{r}i\v{c}ka: Transmittance of
a laser beam through a pair of crossed polarizers; Phys. Lett. A  186 
(1994), 279-82.
 \bibitem{asp}
 A. Aspect, P. Grangier, G. Roger:
Experimental realization of Einstein-Podolsky-Rosen-Bohm
Gedankenexperiment: A new violation of Bell's inequalities; Phys. Rev.
Lett. 49 (1982), 91-4.
 \bibitem{lok}
 M. Lokaj\'{\i}\v{c}ek:  Locality problem,  Bell's inequalities  and
 EPR experiments; submitted to Phys. Rev. Lett.; quant-ph/9808005 (1998).

\end{thebibliography}    }
\end{document}